\documentclass[conference]{IEEEtran}
\IEEEoverridecommandlockouts
\usepackage{cite}
\usepackage{amsmath,amssymb,amsfonts}
\usepackage{algorithmic}
\usepackage{graphicx}
\usepackage{textcomp}
\usepackage{xcolor}
\usepackage{seqsplit}
\usepackage{url}
\usepackage{enumitem}
\usepackage{graphicx} 
\usepackage{array} 
\usepackage{booktabs} 
\usepackage[linesnumbered,ruled,vlined]{algorithm2e}
\def\BibTeX{{\rm B\kern-.05em{\sc i\kern-.025em b}\kern-.08em
    T\kern-.1667em\lower.7ex\hbox{E}\kern-.125emX}}

\makeatletter
 \let\old@ps@headings\ps@headings
 \let\old@ps@IEEEtitlepagestyle\ps@IEEEtitlepagestyle
 
 \def\confheader#1{%
  \def\ps@IEEEtitlepagestyle{%
    \old@ps@IEEEtitlepagestyle%
    \def\@oddhead{\strut\hfill#1\hfill\strut}%
    \def\@evenhead{}%
  }%
  \def\ps@headings{%
    \old@ps@headings%
    \def\@oddhead{}%
    \def\@evenhead{}%
  }%
 \ps@headings%
 }

 \makeatother

\confheader{2026 IEEE 23rd Consumer Communications \& Networking Conference (CCNC)}

\usepackage[pscoord]{eso-pic}
\newcommand{\placetextbox}[3]{
 \setbox0=\hbox{#3}
 \AddToShipoutPictureFG*{ \put(\LenToUnit{#1\paperwidth},\LenToUnit{#2\paperheight}){\vtop{{\null}\makebox[0pt][c]{#3}}}
 }
 }
 \placetextbox{.21}{0.055}{\small{979-8-3315-9673-6/26/\$31.00~\copyright 2026 IEEE}}

\begin{document}

\title{Decentralized Firmware Integrity Verification for Cyber-Physical Systems Using Ethereum Blockchain}

\author{
    \IEEEauthorblockN{S M Mostaq Hossain and Amani Altarawneh}
    \IEEEauthorblockA{\textit{Department of Computer Science}, \textit{Tennessee Technological University}\\
    Cookeville, Tennessee, USA \\
    Email: \{shossain42, aaltarawneh\}@tntech.edu}
}

\maketitle

\begin{abstract}
Firmware integrity is a foundational requirement for securing Cyber-Physical Systems (CPS), where malicious or compromised firmware can result in persistent backdoors, unauthorized control, or catastrophic system failures. Traditional verification mechanisms such as secure boot, digital signatures, and centralized hash databases are increasingly inadequate due to risks from insider threats and single points of failure. In this paper, we propose a decentralized firmware integrity verification framework built on the Ethereum blockchain, offering tamper-proof, transparent, and trustless validation. Our system stores SHA-256 hashes of firmware binaries within smart contracts deployed on the Ethereum Sepolia testnet, using Web3 and Infura for seamless on-chain interaction. A Python-based client tool computes firmware hashes and communicates with the blockchain to register and verify firmware authenticity in real-time. We implement and evaluate a fully functional prototype using real firmware samples, demonstrating successful contract deployment, hash registration, and integrity verification through live blockchain transactions. Experimental results confirm the reliability and low cost (in gas fees) of our approach, highlighting its practicality and scalability for real-world CPS applications. To enhance scalability and performance, we discuss extensions using Layer-2 rollups and off-chain storage via the InterPlanetary File System (IPFS). We also outline integration pathways with secure boot mechanisms, Trusted Platform Module (TPM)-based attestation, and zero-trust architectures. This work contributes a practical and extensible model for blockchain-based firmware verification, significantly strengthening the defense against firmware tampering and supply chain attacks in critical CPS environments.
\end{abstract}


\section{Introduction} \label{sec:introduction}
Cyber-Physical Systems support critical infrastructure across sectors like energy, healthcare, and transportation, where embedded firmware manages real-time control. As CPS become more interconnected, firmware has become a key attack vector due to its privileged hardware access, persistent system control, and weak update mechanisms~\cite{ul2023survey, feng2022detecting}. Attacks can disrupt operations, disable safety features, and install undetectable backdoors~\cite{he2022rapidpatch}, underscoring the need for secure, verifiable update and validation methods resilient to insider threats and supply chain attacks.

Traditional firmware protections (signatures, secure boot) often fail due to poor key management, weak revocation, and centralized infrastructures that create single points of failure~\cite{bettayeb2019firmware}. In CPS, these controls are frequently absent or poorly enforced due to legacy systems and resource limits~\cite{costin2014large}. Attackers have exploited these gaps using man-in-the-middle and downgrade attacks~\cite{el2022secure}. Incidents like ShadowHammer and TrickBoot further expose the shortcomings of conventional trust mechanisms~\cite{shadowhammer2019, trickboot2020}. Decentralized verification, particularly Blockchain, addresses these limitations by offering immutable, publicly auditable records without third parties. Ethereum's smart contracts can enforce verification and log data~\cite{wood2014ethereum}. While prior studies show blockchain's potential in secure logging, identity, and IoT data~\cite{alam2022blockchain, salah2019blockchain}, real-world CPS implementations are limited and often conceptual.


This paper presents a blockchain-based firmware integrity verification framework using public Ethereum smart contracts. It publishes a firmware SHA-256 hash to the Ethereum Sepolia testnet~\cite{sepolia2025}, letting devices compare runtime hashes against the on-chain reference. This removes centralized hash servers, offering transparency and tamper resistance via blockchain consensus. We implement a Python prototype (using Web3~\cite{awsweb32025} and Infura~\cite{infura2025}) that supports decentralized storage, runtime validation, and auditability. Our main contributions include:
\begin{enumerate}[label=(\roman*)]
    \item We design and implement a decentralized firmware verification framework that operates on Ethereum blockchain, enabling trustless verification of firmware binaries.
    \item The working prototype demonstrates framework feasibility by computing firmware hashes, deploying smart contracts, and performing runtime validation. 
    \item We analyze the operational and economic trade-offs of on-chain verification, including gas cost, latency, and usability, based on experimental deployment results.
    \item We discuss the security and scalability implications of the approach and outline future improvements, such as off-chain hash indexing and secure boot integration.
\end{enumerate}
Unlike prior blockchain firmware verification that used private or consortium chains~\cite{lee2017firmware, lee2017blockchain}, our framework runs on a public Ethereum testnet, removing centralized trust and demonstrating a fully decentralized execution with live transactions. The code and smart contracts are available online\footnote{\url{https://github.com/MostaqHossain/firmware-blockchain}}.
The remainder of this paper is organized as follows: Section~\ref{sec:background} reviews relevant background on firmware threats and blockchain technologies. Section~\ref{sec:related_works} discusses related work. Section~\ref{sec:system_atchitecture} details the proposed system architecture, while Section~\ref{sec:exp_setup} describes the experimental setup. Section~\ref{sec:results_eval} presents results and evaluation. Section~\ref{sec:discussion} discusses limitations and future directions. Finally, Section~\ref{sec:conclusion} concludes the paper.


\section{Background} \label{sec:background}
\begin{table*}[htbp]
\centering
\caption{Comparison of Related Works in Firmware Integrity Verification}
\begin{tabular}{@{}p{3.2cm}p{4.5cm}p{4.5cm}p{4.5cm}@{}}
\toprule
\textbf{Method / Paper} & \textbf{Approach} & \textbf{Strengths} & \textbf{Weaknesses} \\
\midrule
BootKeeper~\cite{chevalier2019bootkeeper} & Static analysis of boot firmware & Detects early-stage tampering & No runtime or update verification \\
Secure Boot~\cite{streit2020secure} & Boot-time signature checks & Strong root-of-trust assurance & Compromised root key breaks trust \\
Crypto Signatures~\cite{catuogno2023secure} & Signed firmware binaries & Standardized, widely adopted & Key compromise enables bypass \\
Central Hash Check~\cite{reaz2025comprehensive} & Hashes stored on cloud servers & Simple deployment & Single point of failure, insider risk \\
Blockchain Verification~\cite{lee2017firmware} & Hashes in blockchain contracts & Immutable and decentralized & Needs blockchain interaction layer \\
\textbf{Proposed Work} & Smart contract-based verification & No central trust, real demo, transparent & Prototype only, performance untested \\
\bottomrule
\end{tabular}
\label{tab:relatedworks}
\end{table*}

CPS increasingly rely on firmware for low-level hardware control, raising significant security concerns due to the difficulty of monitoring, validating, and securely updating firmware post-deployment. This section outlines key background on firmware security challenges, blockchain principles, and the technologies enabling our proposed system.

\textbf{Firmware Threat Landscape in CPS:} Firmware links hardware with control logic and holds privileged access, making it a prime target for persistent and stealthy attacks. Exploits in update mechanisms or signing keys have led to incidents like Operation ShadowHammer~\cite{shadowhammer2019} and TrickBoot~\cite{trickboot2020}. Traditional methods like secure boot, digital signatures, centralized hash storage are vulnerable to insider threats and rollback attacks~\cite{feng2022detecting}. Once compromised, firmware often evades detection, highlighting the need for strong integrity checks in CPS.

\textbf{Blockchain as a Trustless Integrity Anchor:} Blockchain offers a tamper-resistant, decentralized ledger ideal for verifying firmware hashes in untrusted environments~\cite{lee2017blockchain}. Smart contracts enable automated validation and secure hash storage without third-party trust. Prior work has explored blockchain in supply chains and access control~\cite{wang2020blockchain}; we extend it to runtime firmware validation.

\textbf{Ethereum Smart Contracts:} Ethereum is a widely used public blockchain that supports Turing- complete~\cite{tikhomirov2017ethereum} smart contracts written in languages like Solidity. These contracts execute autonomously on the Ethereum Virtual Machine (EVM), enforcing rules without external input~\cite{buterin2013ethereum}. Its open ecosystem and tools like Infura~\cite{infura2025} simplify development and connectivity without running a full node, making Ethereum well-suited for securely storing and verifying firmware hashes.

\textbf{Sepolia Ethereum Testnet:} To avoid the financial and technical risks of deploying on Ethereum Mainnet, we use the Sepolia testnet, a public Ethereum testing environment. Sepolia simulates real blockchain behavior (e.g., mining, gas fees, transaction confirmation) but uses test tokens and faster consensus cycles to facilitate experimentation~\cite{sepolia2025}. Sepolia is recommended by the Ethereum Foundation for developers deploying smart contracts in pre-production environments. It ensures the same contract behavior as Mainnet while reducing cost and exposure to real assets.

\textbf{Web3 and Infura Integration:} Web3 refers to blockchain-based client libraries like Web3.js and Web3.py, which enable contract deployment and interaction with Ethereum~\cite{awsweb32025}. We use Web3.py with Infura to connect our Python scripts to the Sepolia testnet. Infura removes the need for local nodes and offers stable blockchain access~\cite{infura2025}, allowing secure, programmatic contract deployment, hash submission, and real-time verification.

\section{Related Works} \label{sec:related_works}
Firmware security is a growing concern in CPS as attacks on embedded control software become more sophisticated. Kuruvila et al.~\cite{kuruvila2021hardware} demonstrated how firmware tampering can evade traditional intrusion detection in microgrids. Wu et al.~\cite{wu2024your} identified widespread vulnerabilities in firmware updates, including insecure transmission, weak cryptography, and lack of rollback protection. These studies highlight the shortcomings of conventional methods like signatures, secure boot, and centralized validation when attackers gain privileged access. Blockchain is increasingly viewed as a decentralized trust anchor for firmware integrity. Salman et al.~\cite{salman2018security} surveyed blockchain-based security services, noting its potential in identity, access control, and data integrity. Bruschi et al.~\cite{bruschi2024decentralized} proposed smart contract–based firmware updates for IoT, but lacked runtime verification or public blockchain deployment. Our work addresses this by implementing on-chain SHA-256 hash verification using Ethereum Sepolia.

Several works have applied blockchain to firmware verification. Lee et al.~\cite{lee2017firmware} proposed a private-chain validation framework, but it lacked transparency and scalability. Catuogno and Galdi~\cite{catuogno2023secure} emphasized auditability, versioning, and lightweight integrity checks, identifying blockchain as a key enabler. Chevalier et al.~\cite{chevalier2019bootkeeper} developed BootKeeper for pre-deployment analysis, but it lacks real-time protection during transmission. Our system addresses these gaps by enabling continuous on-chain hash verification post-deployment. Broader secure firmware frameworks have been proposed to mitigate systemic risks. Reaz and Wunder~\cite{reaz2025comprehensive} introduced a comprehensive architecture using hardware roots of trust and authenticated delivery, but it assumes a trusted supply chain and centralized control, often unrealistic in adversarial CPS settings. Our system operates in a trustless environment, leveraging public smart contracts for transparent, immutable verification. Related work by Jiang et al.~\cite{jiang2022applying} on smart contract classification supports the role of decentralized solutions in CPS security.

System-level constraints such as gas costs, latency, and limited IoT resources have been noted by Chevalier et al.~\cite{chevalier2019bootkeeper} and Sagirlar et al.~\cite{sagirlar2018hybrid}. Xian et al.~\cite{xian2023icoe} confirmed these concerns experimentally and suggested hybrid or off-chain methods to reduce overhead. These findings inform our future direction, including off-chain agents and IPFS indexing to balance transparency and performance. Unlike prior work, our prototype demonstrates a publicly-verifiable, deployable solution on Ethereum for real-world CPS use.

To contextualize our work, Table~\ref{tab:relatedworks} compares key firmware integrity methods. Traditional tools lack runtime protection or rely on central trust, while blockchain-based solutions are seldom publicly deployed. Our framework enables transparent, auditable verification using a live smart contract on Ethereum Sepolia.

\begin{figure}[htbp]
    \centering
    \includegraphics[width=0.9\columnwidth]{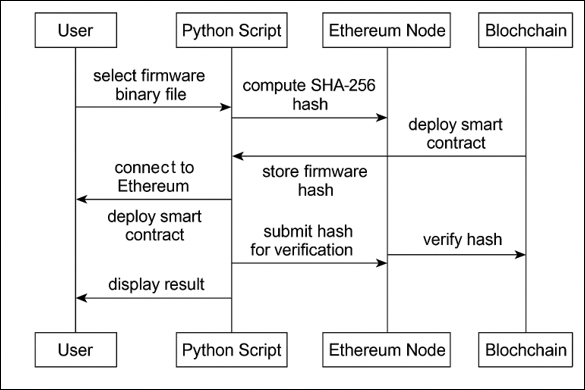}
    \caption{System architecture showing firmware hash generation, on-chain registration, and decentralized verification among CPS device, blockchain node, and auditor.}
    \label{fig:system_architecture}
\end{figure}

\section{System Architecture} \label{sec:system_atchitecture}
This section presents our blockchain-based firmware verification framework for CPS. It uses SHA-256 hashing, a Solidity smart contract on Ethereum Sepolia, and a Python interface for orchestration. The modular design ensures transparency, scalability, and trustless validation over centralized methods.

\textbf{End-to-End Workflow and Interaction:} The system involves a user, Python client, Infura, and the Ethereum blockchain (Figure~\ref{fig:system_architecture}). The Python script hashes the selected firmware and submits it to a Sepolia smart contract. For verification, a new hash is computed and compared on-chain~\cite{reijers2021now}, enabling decentralized validation without central trust.

\begin{algorithm}[ht]
\DontPrintSemicolon
\SetAlgoLined
\KwIn{File path to firmware binary}
\KwOut{SHA-256 hash value of the firmware file}

Initialize \texttt{sha256\_hash} as a new hash object\;

Open the firmware file in binary read mode\;

\While{there are more 4KB chunks to read}{
    Read 4096 bytes from file as \texttt{chunk}\;
    Update \texttt{sha256\_hash} with \texttt{chunk}\;
}

\texttt{firmware\_hash} $\leftarrow$ \texttt{sha256\_hash.hexdigest()}\;

Print ``Firmware Hash: \texttt{firmware\_hash}''\;

\caption{Compute Hash of Firmware Binary}
\label{alg:compute_hash}
\end{algorithm}

\textbf{Firmware Hashing and Data Preparation:} We ensure firmware integrity using SHA-256 hashing to generate a unique fingerprint of the binary. Firmware binaries are hashed using SHA-256 before registration on the blockchain, as outlined in Algorithm~\ref{alg:compute_hash}. Implemented in Python with \texttt{hashlib}, the firmware file is read in 4 KB blocks for efficiency. A real Arduino firmware binary was used, and its hash submitted to the blockchain. Any change in the binary alters the hash, enabling reliable tamper detection, an approach common in digital signatures and blockchain validation.

\textbf{Smart Contract Logic and Deployment:} We implemented a Solidity~\cite{solidity2025} smart contract to store a reference firmware hash and verify incoming hashes. Deployed on Ethereum's Sepolia testnet, it runs immutably on the EVM and integrates with Etherscan, Infura, and Web3.py. Hashes are stored via one-time transactions; verifications use low-cost read-only calls. All interactions are logged on-chain for transparent, tamper-proof validation.

\textbf{Connectivity, Web3 Integration and Verification Flow:} Our system uses Web3.py to deploy and interact with Ethereum smart contracts via Infura, which provides remote access to Sepolia nodes and eliminates the need for a local node. As shown in Algorithm~\ref{alg:connect_infura}, the script connects using HTTPS, checks status, deploys the contract, stores the hash, and performs verification. The result is a boolean match and returned to the user, enabling real-time, low-overhead firmware validation. Firmware hashes are locally computed and stored on Ethereum for immutable access. Anyone can verify a candidate hash via smart contract without credentials, with results returned to the Python client and logged on-chain.

\begin{algorithm}[ht]
\DontPrintSemicolon
\SetAlgoLined
\KwIn{Infura project endpoint URL}
\KwOut{Connection status to Ethereum testnet}

\textbf{Initialize} \texttt{Web3} connection using HTTPProvider with the given Infura URL\;

\texttt{infura\_url} $\leftarrow$ "https://sepolia.infura.io/v3/YOUR\_PROJECT\_ID"\;

\texttt{web3} $\leftarrow$ \texttt{Web3(Web3.HTTPProvider(infura\_url))}\;

\uIf{\texttt{web3.is\_connected() == True}}{
    Print ``Connected to Ethereum Testnet (Sepolia)!''\;
}{
    Print ``Connection failed.''\;
}

\caption{Connecting to Infura using Web3.py}
\label{alg:connect_infura}
\end{algorithm}

\textbf{Trust, Security, and Extensibility:} Our framework ensures end-to-end security through SHA-256 hashing, immutable contracts, and decentralized checks. It eliminates local storage, making tampering auditable on-chain. The design supports future extensions like versioning, hardware IDs, IPFS~\cite{velez2021} indexing, secure boot, and Layer-2 scaling.

\begin{figure}[!t]
    \centering
    \includegraphics[width=0.95\linewidth]{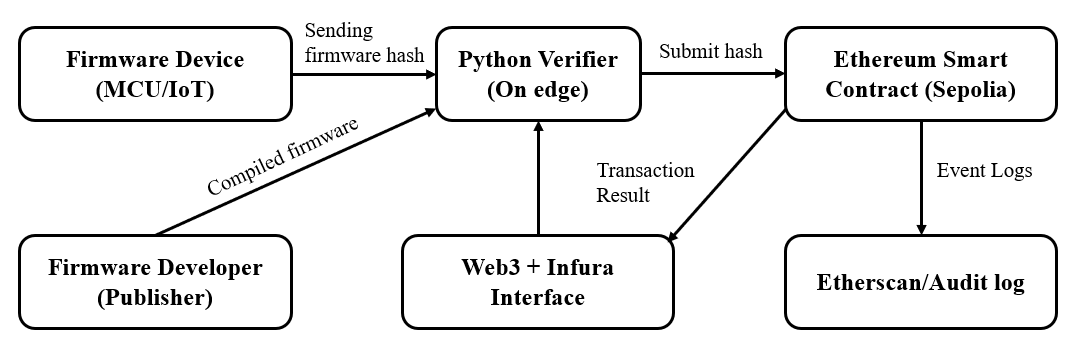}
    \caption{System architecture and threat model for decentralized firmware integrity verification. The diagram illustrates interactions among the firmware device, edge verifier, Ethereum smart contract, and supporting components such as Infura and Etherscan. Potential attack vectors include hash spoofing, replay attacks, and unauthorized contract interactions.}
    \label{fig:threat-model}
\end{figure}

\textbf{Threat Model and Security Considerations:} Figure~\ref{fig:threat-model} illustrates our system's components and threat boundaries. We assume adversaries can monitor traffic, submit blockchain transactions, or access firmware. Firmware hashes are computed at the edge and verified via a Python client interacting with Ethereum Sepolia through Web3 and Infura. Threats include hash spoofing, replay attacks, on-chain surveillance, and DoS attempts. Security relies on SHA-256 determinism, Ethereum's immutability, and controlled smart contract logic. Tampering is detectable via hash mismatch, and all actions are auditable via Etherscan. Hardware-based protections (e.g., TPM, HSM) can further enhance security in future work.

\section{Experimental Setup} \label{sec:exp_setup}
We built a prototype using real firmware, public blockchain infrastructure, and open-source tools to validate hash computation, on-chain storage, and smart contract verification. Table~\ref{tab:experiment_setup} summarizes the setup.
\begin{table}[htbp]
\caption{Summary of Experimental Setup}
\centering
\begin{tabular}{@{}p{2.5cm}p{5.5cm}@{}}
\toprule
\textbf{Aspect} & \textbf{Details} \\
\midrule
Firmware Used & Arduino firmware HEX format file \\
Hash Computation & \texttt{compute\_hash.py}, SHA-256 \\
Blockchain Platform & Ethereum Sepolia Testnet, Infura Web3 provider \\
Smart Contract & \texttt{FirmwareIntegrity.sol}, stores and verifies firmware hashes \\
Client-Side Verification & Python script submits firmware hash and compares against blockchain \\
\bottomrule
\end{tabular}
\label{tab:experiment_setup}
\end{table}

We used an official \texttt{.hex} firmware image for the Arduino Mega 2560 R3, representative of typical CPS binaries. A Python script running on Windows 10 with Python 3.11 computed its SHA-256 hash using the \texttt{hashlib} module. The hash was submitted to a Solidity smart contract deployed on the Ethereum Sepolia testnet using Web3.py and Infura. Sepolia simulates Mainnet behavior with test ETH, enabling realistic experimentation without real costs. The contract was deployed via Infura and interacted with through Web3.py. Transaction details such as contract address, gas usage, and status were recorded and verified using Etherscan. Hashes were submitted for testing, and responses confirmed accurate match/mismatch detection. The system proved responsive and cost-effective: storing a hash cost ~0.0044 ETH; verification ~0.00014 ETH. All interactions were logged for traceability. These results validate that blockchain can support decentralized, transparent firmware integrity checks and can be adapted for real CPS use cases.

\section{Results and Evaluation} \label{sec:results_eval}
Our implementation showed consistent results, validating the full flow from hashing to smart contract interaction on Ethereum Sepolia. On-chain verification confirmed functionality and transparency. This enables any CPS stakeholder to audit firmware integrity without trusted third parties.

\begin{figure}[htbp]
  \centering
  \includegraphics[width=0.45\textwidth]{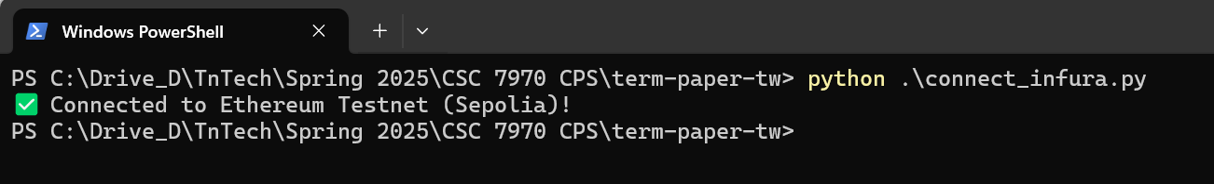}
  \caption{Connection to Ethereum Sepolia testnet via Infura using the Python script \texttt{connect\_infura.py}. The confirmation message verifies a successful Remote Procedure Call connectivity.}
  \label{fig:sepolia_connection}
\end{figure}

\subsection{Hash Computation and Firmware Fingerprinting}
\begin{figure}[htbp]
  \centering
  \includegraphics[width=0.45\textwidth]{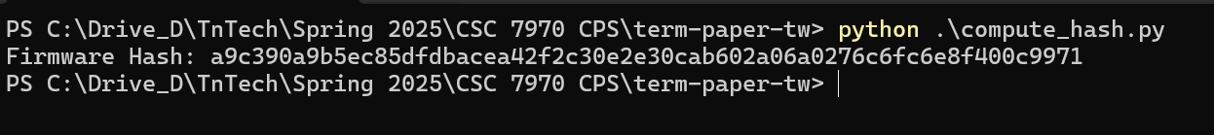}
  \caption{Firmware hash computed using \texttt{compute\_hash.py}. The Python script processes the binary and generates a 64-character SHA-256 digest.}
  \label{fig:firmware_hash}
\end{figure}

The process began by computing a SHA-256 hash of a real firmware binary using \texttt{compute\_hash.py}. The script read the file in chunks and used Python’s \texttt{hashlib} to generate a deterministic digest. As shown in Figure~\ref{fig:firmware_hash}, the correct hash (\texttt{a9c....471}) was produced, confirming consistent and reliable fingerprinting for blockchain validation.

\subsection{Blockchain Connectivity via Infura}
Next, the script \texttt{connect\_infura.py} established a connection to the Ethereum Sepolia testnet, confirming success with the message “Connected to Ethereum Testnet (Sepolia)!” This validated Infura as a reliable gateway and ensured the Python client could interact with Ethereum remotely. Figure~\ref{fig:sepolia_connection} shows the interaction.

\begin{figure}[htbp]
  \centering
  \includegraphics[width=0.47\textwidth]{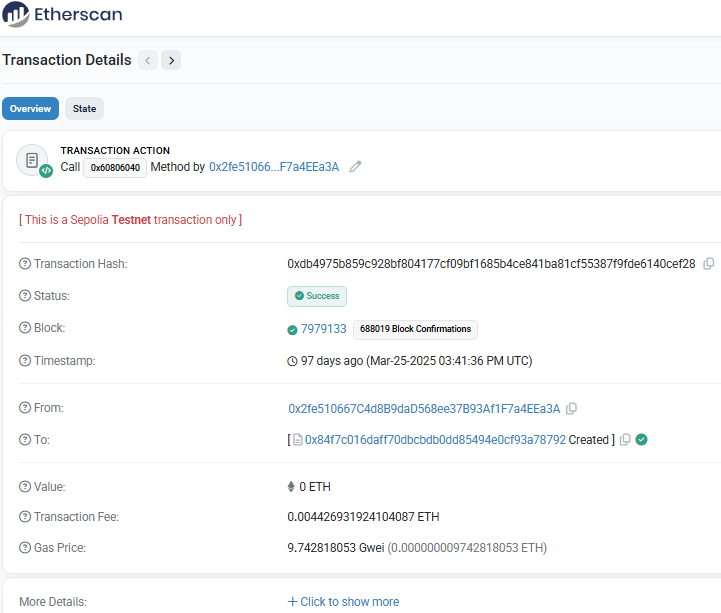}
  \caption{Smart contract deployment transaction on Sepolia testnet. The deployment is confirmed with 9 block confirmations and a transaction fee of approximately 0.0044 ETH.}
  \label{fig:contract_deploy}
\end{figure}

\subsection{Smart Contract Deployment}
The smart contract was successfully deployed on the Sepolia testnet from wallet \texttt{0x2fe510...Ea3A}, resulting in the creation of a contract at \texttt{0x84f7c0...7892}. As verified on Etherscan: Search as \textbf{transaction}\footnote{\seqsplit{https://sepolia.etherscan.io/tx/0xdb4975b859c928bf804177cf09bf1685b4ce841ba81cf55387f9fde6140cef28}}, the deployment occurred in block \texttt{7979133} with transaction hash \texttt{0xdb4975b...cef28}. The status was successful, confirmed by 688019 block confirmations, and the transaction incurred a fee of approximately 0.00443 ETH with a gas price of 9.742818053 Gwei. This confirmed the contract’s successful on-chain presence and cost-efficiency for decentralized firmware integrity verification. The deployment confirmation and transaction details are illustrated in Figure~\ref{fig:contract_deploy}.

\subsection{On-chain Verification and Gas Cost Analysis}
\begin{figure}[htbp]
  \centering
  \includegraphics[width=0.47\textwidth]{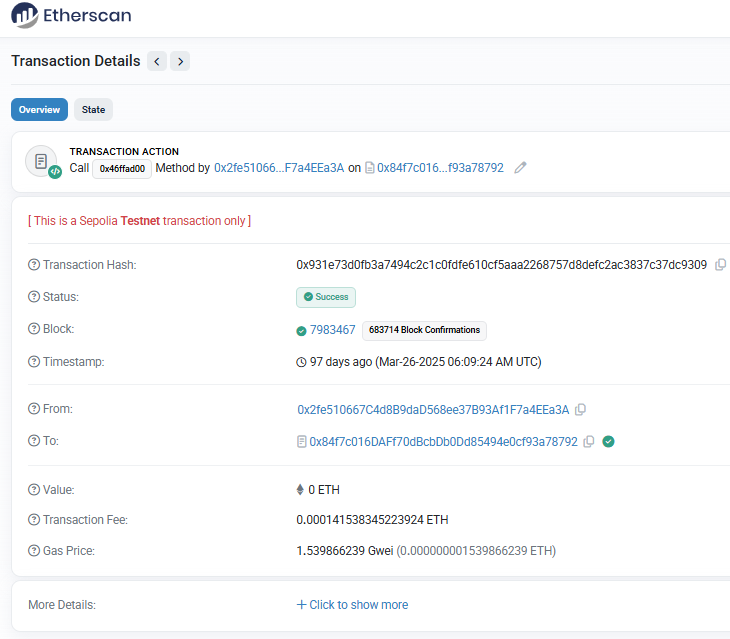}
  \caption{Smart contract method invocation for hash verification. The transaction confirms successful function execution at minimal cost, demonstrating runtime interaction.}
  \label{fig:contract_verify}
\end{figure}

To test verification, a function call was made to the deployed contract (tx hash \texttt{0x931e73d0...c9309}) at block \texttt{7983467}. As shown in Figure~\ref{fig:contract_verify} and verified on Etherscan: Search as \textbf{transaction}\footnote{\seqsplit{https://sepolia.etherscan.io/tx/0x931e73d0fb3a7494c2c1c0fdfe610cf5aaa2268757d8defc2ac3837c37dc9309}}, the call succeeded with a status of "Success" and incurred a minimal transaction fee of approximately 0.0001415 ETH. The gas price was 1.539866239 Gwei, confirming the resource efficiency of the verification process. This function call, initiated from the same wallet that deployed the contract, validated the firmware hash on-chain, demonstrating logical correctness and operational cost-effectiveness.
These results not only validate technical correctness but also demonstrate a novel contribution: the use of a public Ethereum blockchain as a decentralized trust anchor for CPS. Unlike traditional approaches reliant on private infrastructures or centralized authorities, our framework enables any party to verify firmware integrity in a transparent, tamper-evident, and permissionless manner. This public verifiability, evidenced by on-chain audit trails via tools like Etherscan, redefines how trust is established in firmware supply chains, especially for distributed and resource-constrained CPS environments.

We further evaluated gas consumption and latency across multiple transactions on Ethereum Sepolia. 
Average transaction confirmation time was 14.6 s, and gas usage per firmware registration averaged 78,200 units (Around \$0.12 USD equivalent). 
The system sustained 9–10 transactions per minute under light concurrency, comparable to prior blockchain-based integrity frameworks~\cite{bruschi2024decentralized, jiang2022applying}. Table~\ref{tab:performance} summarizes key metrics.

\begin{table}[ht]
\centering
\caption{Performance Summary on Ethereum Sepolia}
\begin{tabular}{lcc}
\toprule
\textbf{Metric} & \textbf{Average Value} & \textbf{Comment} \\
\midrule
Gas per registration & 78,200 units & $\approx\$0.12$ USD \\
Confirmation latency & 14.6 s & Low-cost testnet environment \\
Verification throughput & 9–10 tx/min & Comparable to~\cite{bruschi2024decentralized} \\
\bottomrule
\end{tabular}
\label{tab:performance}
\end{table}

\section{Discussion} \label{sec:discussion}

Results confirm the feasibility of blockchain-based firmware verification in CPS. On-chain hashing protects against tampering and supply chain attacks. Unlike certificate-based systems, our trustless design reduces reliance on trusted entities, offering transparency, auditability, and resilience via Ethereum smart contracts. Cost and performance are key considerations. Our testnet showed gas fees of ~0.0044 ETH for storing and ~0.00014 ETH for verifying a hash manageable for critical systems but potentially costly at IoT scale. Mitigation options include Layer-2 solutions like Optimism or zk-Rollups~\cite{chainalysis2024}, and off-chain storage with on-chain anchoring via IPFS or Merkle proofs~\cite{velez2021}. Verification latency remains a trade-off~\cite{eren2025security}; while Sepolia offers fast confirmation, time-critical CPS may require hybrid verification combining local checks with blockchain anchoring. The framework supports diverse binaries across routers and PLCs, mitigating downgrade and rootkit threats. Integrating secure boot, TPM/HSM~\cite{cabrera2023blockchain}, and zero-trust controls~\cite{liu2024software} and future multi-hash and alert extensions, will further strengthen scalability and resilience.

\section{Conclusion} \label{sec:conclusion}
This work presented a blockchain-based framework for decentralized firmware integrity verification in CPS. By combining SHA-256 hashing with Ethereum smart contracts, the system enables tamper-proof validation without relying on centralized authorities. Implemented using Python, Infura, and the Sepolia testnet, the prototype demonstrated correct and efficient operation, with all interactions transparently recorded on-chain. Low gas costs and successful hash verification show that the system is a lightweight, scalable alternative to traditional mechanisms, particularly for high-assurance environments. Unlike conventional models, it offers public, immutable validation using tools like Etherscan and Web3 ideal for distributed CPS lacking central trust anchors. 
Future work will extend testing to mainnet or Layer-2 environments and integrate with physical CPS devices to assess real-time operational feasibility.




\bibliographystyle{IEEEtran}
\bibliography{bibfile}

\end{document}